\documentstyle[aps,prl]{revtex}
\begin{document}
\bibliographystyle{prsty}
\draft

\title{Band-by-band decompositions of the Born effective charges}
\author{Philippe Ghosez$^1$ and Xavier Gonze$^2$}

\address{$^1$ D\'epartement de Physique, Universit\'e de Li\`ege,
B-5, B-4000 Sart-Tilman, Belgium}
\address{$^2$ Unit\'e de Physico-Chimie et de Physique des Mat\'eriaux,
Universit\'e Catholique de Louvain,\\
1 Place Croix du Sud, B-1348 Louvain-la-Neuve, Belgium}
\date{\today}
\maketitle
\begin{abstract}
The Born effective charge, $Z^*$, that describes the polarization
created by collective atomic displacements, can be computed
from first-principles following different techniques. We establish the
connections existing between these different formulations, and
analyze the related band-by-band decompositions. We show
that unlike for the full $Z^*$, the different band-by-band values are not
equal, and emphasize that one of them has a natural physical meaning
in terms of Wannier functions.
\end{abstract}

\setcounter{page}{1}

\section{Introduction}

The Born effective charge~\cite{Born33} ($Z^*$) is the equivalent,
for crystalline insulating solid, of the atomic polar tensor introduced
for isolated molecules~\cite{Biarge61}. It is a dynamical charge related
to the macroscopic
polarization induced by the collective displacements of nuclei belonging
to a given sublattice. In the study of the lattice dynamics of insulating
crystals, it is considered as a fundamental quantity, because it
governs the amplitude of the long-range Coulomb interaction between
nuclei, and the splitting between longitudinal (LO) and transverse (TO)
optic phonon modes.

In simple materials, like $A^N B^{8-N}$ binary crystals~\cite{Phillips70},
the phonon eigenvectors are imposed by symmetry. Infra-red measurements of the
splitting between LO and TO modes allows an accurate estimation of $|Z^*|^2/
\epsilon_{\infty}$ and offers therefore an unambiguous way to extract
the amplitude of $Z^*$ from the experiment. However, in more complex materials
like ABO$_3$ compounds, LO and TO mode eigenvectors are not necessarily
equivalent. The determination of $Z^*$ from the experimental data is
consequently not straightforward and requires the use of some
approximations~\cite{Axe67}. For such compounds, the development of
theoretical methods giving direct access to $Z^*$ acquires therefore a
specific interest.

Conventionally, the Born effective charge tensor $Z^*_{\kappa,\alpha\beta}$
of nuclei belonging to the sublattice $\kappa$ is defined as
the coefficient of proportionality relating, under the condition of zero
macroscopic electric field, the change in macroscopic polarization
${\cal P}_\beta$ along the direction $\beta$ and the collective nuclear
displacements of atoms $\kappa$ along direction $\alpha$, times the unit
cell volume $\Omega_0$~:
\begin{equation}
\label{EqZ.P}
Z^*_{\kappa,\alpha\beta}
=
\Omega_0 \;
   {\left .
    \frac{\partial{\cal P}_{\beta}}
            {\partial  \tau_{\kappa\alpha}}
    \right |}_{{\cal E}=0}.
\end{equation}
However, a thermodynamical equality relates the macroscopic polarization
to a derivative of the electric enthalpy ${\tilde E}$ and another
relationship connects the forces $F_{\kappa}$ on the nuclei $\kappa$ to a
derivative of the electric enthalpy so that $Z^{*}$ can be alternatively
defined as follows:
\begin{equation}
\label{EqZ.2}
Z^*_{\kappa,\alpha\beta}
=
- \frac{\partial^2 {\tilde E}}
            {\partial {\cal E}_\beta \partial \tau_{\kappa\alpha}}
=
     {\left .
     \frac{\partial F_{\kappa,\alpha}}
             {\partial {\cal E}_\beta}
     \right |}_{\tau_{\kappa \alpha}=0}.
\end{equation}
 From these relationships, Z$^{*}$ can therefore be thought either as (i)
the change of polarization induced by the collective displacements of
atoms $\kappa$, under zero field (ii) a mixed second derivative of the
electric enthalpy or (iii) the derivative of the force induced on a
nucleus $\kappa$ by an homogeneous effective electric field
${\cal E}_\beta$, at zero atomic displacements.

The three definitions are formally equivalent. However, each of them can
lead to different algorithms for the computation of $Z^{*}$ from
first-principles. Among the approaches which are the most widely used, a first
powerful and systematic procedure was introduced by Baroni, Giannozzi
and Testa~\cite{Baroni87}, who suggested to determine $Z^*$ from a linear
response formalism grounded on a Sternheimer equation. A different algorithm,
based on a variational principle, was later reported by Gonze, Allan and
Teter~\cite{Gonze92}, yielding a new alternative expression for $Z^*$.
Thanks to progress in the theory of the macroscopic polarization,
$Z^*$ is also directly accessible from finite difference of
polarization~\cite{KingSmith93}. The first two algorithms were exclusively
implemented within the density functional formalism (DFT) while the last
one also allowed calculations of changes
in polarization within different other one-electron schemes (Hartree-Fock
method~\cite{DallOlio97}, model GW approximations to many-body
theory~\cite{Massidda95,Gonze97c}, Harrison tight-binding
model~\cite{Bennetto96}) and the Hubbard tight-binding model~\cite{Resta95}.

Accurate predictions of the Born effective charges have been reported
for a large variety of materials. In some of these
studies~\cite{Massidda95,Ghosez95a,Ghosez95b,Posternak97,Ghosez98,Marzari98},
the decomposition of $Z^{*}$ in individual contributions from separate groups
of occupied bands appeared as a powerful tool to identify the microscopic
mechanisms monitoring its amplitude.  However, the physical
interpretation of these decompositions was never explicitely discussed.
Moreover, unlike the total $Z^{*}$, contributions from individual groups of
bands are not uniquely defined.

In the present paper, we aim at presenting the links between the
theoretical frameworks used nowadays to compute $Z^{*}$, and
at deducing from this comparison the correct way to develop a band-by-band
analysis. We demonstrate that the natural decompositions arising from
Eq.~(\ref{EqZ.P}) and Eq.~(\ref{EqZ.2}) {\it differs}, although independent
approches (linear response, Berry phase approach, electronic Wannier
functions) to Eq.~(\ref{EqZ.2}) are strictly {\it equivalent} provided
the phase of the wavefunctions are correctly chosen.

The paper is organized as follows. In Section II, we describe the
mathematical links existing between the different expressions that can be
used to determine the global value of $Z_{\kappa,\alpha\beta}$ within the
density functional formalism without yet referring to band-by-band
decompositions. We adopt the notations of Ref.~\cite{Gonze97a,Gonze97b}.
In Section III, we discuss how contributions
from isolated sets of bands can be separated from each others. We
identify different expressions and discuss their meaning in terms of
electronic Wannier functions. In Section IV, we illustrate our results
on a numerical example, emphasizing that independent decompositions
yield in practice radically different values. Finally, in Section V,
we conclude on the physically correct way of performing band-by-band
decomposition of $Z^{*}$.

\section{Different formulations of Z*}


For practical purposes, in what follows, the Born effective charge will be
decomposed into two contributions~:
\begin{equation}
\label{Eq5.2}
Z^*_{\kappa,\alpha\beta}
= Z_\kappa \delta_{\alpha\beta} + Z_{\kappa,\alpha\beta}^{el}.
\end{equation}
The first term, $Z_\kappa$, is the charge of the nuclei (or
pseudo-ion, in case of pseudopotential calculations), and can be trivially
assigned. The second, $Z_{\kappa,\alpha\beta}^{el}$, is the contribution
due to the electrons.

\subsection{First derivative of the polarization}

A first straightforward approach for the determination of
$Z^*_{\kappa}$ consists in computing the difference of macroscopic
polarization between a reference state, and a state where the atoms
belonging to the sublattice $\kappa$ have been displaced by a small but finite
distance $\Delta \tau_{\kappa, \alpha}$. The electronic contribution to
$Z^{*}$ can be obtained as~:
\begin{equation}
\label{EqPf}
Z^{el}_{\kappa,\alpha\beta}
=
\Omega_o
\lim_{\Delta \tau_{\kappa, \alpha} \rightarrow 0} \;
\frac{\Delta {\cal P}^{el}_{\beta}}{\Delta \tau_{\kappa, \alpha}}
\end{equation}
In periodic systems, the change in electronic polarization in zero field
can be computed from the King-Smith and Vanderbilt formula~\cite{KingSmith93}~:
\begin{equation}
\label{Pol}
{\cal P}_{\beta}^{el}= - \frac{1}{(2 \pi)^3} \; i
    \sum_{n}^{occ} s
        \int_{BZ}
           \langle
    u_{n{\bf k}} | \frac{\partial}{\partial k_{\beta}} | u_{n{\bf k}}
           \rangle
       \; \;  d{\bf k}
\end{equation}
where $s$ is the occupation number of states in the valence bands ($s=2$ in
spin-degenerate system) and $u_{n{\bf k}}$ is the periodic part of the Bloch
functions. Taken independently, the matrix elements of the previous 
equation are
ill-defined because the phase of the wavefunctions at a given wavevector
of the Brillouin zone is arbitrary, and thus unrelated with the phases at
neighbouring {\bf k} points. However, the {\it integral} of the right-hand side
is a well-defined quantity, which takes the form of a Berry phase of band $n$,
as discussed by Zak~\cite{Zak89}.

The King-Smith and Vanderbilt definition is valid  {\it only} under
the constraint that the wavefunctions fulfill the {\em periodic gauge}
condition. This means that the periodic part of Bloch functions must satisfy
\begin{equation}
\label{Eq.PG}
u_{n {\bf k}}({\bf r})= e^{i {\bf G.r}} \;
u_{n {\bf k+G}}({\bf r}) ,
\end{equation}
This condition does not fix unambiguously the phase of the wavefunctions at
a given {\bf k}-point (even not at neighbouring {\bf k}-points) but it imposes
a constraint for wavefunctions at distant wavevectors. It defines a topology
in {\bf k}-space, within which the polarization takes the convenient form of
a Berry phase.

When working within one-electron schemes (DFT, Hartree-Fock, ...), a
second choice of phase is present at {\it another} level.
For the ground-state, the Lagrange multiplier method applied to the
minimization of the Hohenberg and Kohn fonctional under orthonormalization
conditions on the wavefunctions~\cite{Gonze95b}, gives the following
equations~:
\begin{equation}
\label{eq.s}
H_{\bf k} | u_{m{\bf k}} \rangle =
\sum_{n}^{occ} \Lambda_{mn,{\bf k}} \;
     | u_{n{\bf k}} \rangle
\end{equation}
This condition, associated with the minimisation of the Hohenberg
and Kohn energy functional, means that the application of the
Hamiltonian to a given wavefunction generates a vector which must
stay within the Hilbert space defined by the set of $u_{n{\bf k}}$
wavefunctions. We observe that a unitary transformation between the
wavefunctions will leave that Hilbert space invariant, and
Eq.~(\ref{eq.s}) will remain satisfied provided the matrix of
Lagrange multiplier $\Lambda_{mn,{\bf k}}$ is transformed accordingly.
In order to build Kohn-Sham band structures, the unitary transform
is implicitely chosen such as to guarantee
\begin{equation}
\label{eq.diag}
\Lambda_{mn,{\bf k}}=
\delta_{mn} \; \epsilon_{m,{\bf k}}
\end{equation}
in which case
$\epsilon_{m,{\bf k}}$ correspond to the eigenvalues of the
Kohn-Sham Hamiltonian and the associated functions $u_{d,m{\bf k}}$ are the
Kohn-Sham orbitals. This choice is called the {\em diagonal gauge} condition.
All along this work, it will be emphasized by a ``$d$'' subscript.

We note that the periodic gauge condition connect wavefunctions at
different {\bf
k}-points, while the diagonal gauge condition fixes wavefunctions at a
given {\bf
k}-point. The choice defined by Eq.~(\ref{eq.diag}) is {\it not} mandatory,
and the computation of the total energy, the density, or the
Berry phase (Eq.~(\ref{Pol})) will give the {\it same} value independently
of the fulfillment of Eq.~(\ref{eq.diag}). The diagonal gauge is
the natural choice for the ground-state wavefunctions while, as it will be
discussed later, another choice is usually
preferred for the change in wavefunctions in linear-response calculations.


Instead of approximating Eq. (\ref{EqZ.P}) from finite differences, it can
be computed directly.  The combination of Eqs.~(\ref{EqZ.P}),
(\ref{Eq5.2}) and
(\ref{Pol})  gives~:

\begin{eqnarray}
Z^{el}_{\kappa, \alpha \beta} =
  - \frac{\Omega_o}{(2 \pi)^3} \; i
    \sum_{m}^{occ} s
        \int_{BZ}
   [
           \langle
              \frac{\partial u_{n{\bf k}}}{\partial \tau_{\kappa, \alpha}}
  |
              \frac{\partial u_{n{\bf k}}}{\partial { k}_{\beta}}
           \rangle
  +
           \langle
             u_{n{\bf k}}
  |
             \frac{\partial}{\partial {k}_{\beta}}
  |
             \frac{\partial u_{n{\bf k}}}{\partial \tau_{\kappa, \alpha}}
           \rangle
  ]
        d{\bf k}
\end{eqnarray}
where the second expectation value can be worked out~:
\begin{eqnarray}
&&   \int_{BZ}
           \langle
             u_{n{\bf k}}
  |
             \frac{\partial}{\partial  { k}_{\beta}}
  |
              \frac{\partial u_{n{\bf k}}}{\partial \tau_{\kappa, \alpha}}
           \rangle
     d{\bf k}
=
      \int_{BZ}
  [
      \frac{\partial}{\partial  { k}_{\beta}}
           \langle
             u_{n{\bf k}}
  |
             \frac{\partial u_{n{\bf k}}}{\partial  \tau_{\kappa, \alpha}}
           \rangle
  -
           \langle
             \frac{\partial  u_{n{\bf k}}}{\partial  { k}_{\beta}}
  |
             \frac{\partial u_{n{\bf k}}}{\partial  \tau_{\kappa, \alpha}}
           \rangle
]
        d{\bf k}.
\end{eqnarray}
In the last expression, the first term of the right-hand side
is the gradient of a periodic quantity
integrated over the Brillouin zone. Within any periodic gauge,
its contribution will be zero. Using the time-reversal symmetry,
we arrive therefore at the final expression~:
\begin{eqnarray}
\label{Z*B}
Z^{el}_{\kappa, \alpha \beta}
=  -2  \frac{\Omega_o}{(2 \pi)^3} \; i
    \sum_{n}^{occ}  s
        \int_{BZ}
        \langle
              \frac{\partial u_{n{\bf k}}}{\partial \tau_{\kappa, \alpha}}
  |
              \frac{\partial u_{n{\bf k}}}{\partial { k}_{\beta}}
       \rangle
       d{\bf k}
\end{eqnarray}

The first-derivatives of the wave functions, ${\partial u_{n{\bf k}}/{\partial
\tau_{\kappa,\alpha}} }$ and ${\partial u_{n{\bf k}}/{\partial {k}_{\beta}}}$,
appearing in this
expression, can be computed by linear-response techniques either by
solving a first-order Sternheimer equation~\cite{Baroni87,Giannozzi91} or by
the direct minimization of a variational expression as described in
Ref.~\cite{Gonze92,Gonze97a}.

We note that the choice of gauge will influence the value of the
first-derivative of $u_{n {\bf k}}$, although the integrated quantity
$Z^{el}_{\kappa, \alpha \beta}$ must remain independent of this choice (in any
periodic gauge). Usually, the following choice is preferred in linear-response
calculations~:
\begin{eqnarray}
        \langle
             {\left .
             \frac{\partial u_{n{\bf k}}}{\partial \lambda}
             \right |}_p
  |
              u_{m{\bf k}}
       \rangle
  = 0
\end{eqnarray}
for $m$ and $n$ labelling occupied states, and $\lambda$
representing either the derivative with respect to the wavevector
or to atomic displacements. As emphasized by the ``$p$'' subscript, this
condition defines what is called the {\em parallel gauge} and insures that the
changes in the occupied wavefunctions are orthogonal to the space of the
ground-state occupied wavefunctions. This projection on the conduction bands is
not reproduced within the diagonal gauge defined by the generalization of
Eq.~(\ref{eq.s}-\ref{eq.diag}) at the first order of perturbation, as
elaborated in Ref.~\cite{Gonze95b}.

\subsection{Mixed second derivatrive of the electric enthalpy}

The Born effective charge also appears as a mixed second derivative of 
the electric enthalpy. Therefore, as reported in Eq.~(41) of 
Ref.~\cite{Gonze97b}, $Z^{el}_{\kappa,\alpha \beta}$ can be alternatively 
formulated in terms of a {\it stationary} expression, involving the 
first-order derivative of the wavefunctions with respect to a collective 
displacement of atoms of the sublattice $\kappa$ and the first-order 
derivatives of the wavefunctions with respect to an electric
field and to their wavevector~\cite{Notation}~.


The mathematical equivalence between Eq.~(\ref{Z*B}) and Eq.~(41) of
Ref.~\cite{Gonze97b} is a consequence of the interchange theorem but 
can be directly highlighted from the stationary character of the latter. 
Indeed, as the error on $Z^*$ is proportional to the product of the 
errors on the first-order change in wavefunctions, if
${\partial u_{n{\bf k}}}/{\partial \tau_{\kappa, \alpha}}$ was known
perfectly, a correct estimation of $Z_{\kappa,\alpha\beta}^{el}$ should be
obtained independently of the knowledge of ${\partial u_{n{\bf 
k}}}/{\partial {\cal
E}_\beta}$. Putting therefore to zero
${\partial u_{n{\bf k}}}/{\partial {\cal E}_\beta}$ and the corresponding
density changes in Eq.~(41) of Ref.~\cite{Gonze97b}, most of the terms 
cancel out and we recover
Eq. (\ref{Z*B}), which evaluated for the exact ${\partial u_{n{\bf 
k}}}/{\partial
\tau_{\kappa, \alpha}}$ must still correspond to a valid expression
for $Z^*$.

\subsection{First derivative of the atomic force}

By the same token as above, we can choose alternatively for
${\partial u_{n{\bf k}}}/{\partial \tau_{\kappa, \alpha}}$ and
the associated density derivative to vanish
in Eq.~(41) of Ref.~\cite{Gonze97b}, and we still obtain a valid 
expression for $Z^*$~:
\begin{eqnarray}
\label{Eq5.9}
Z_{\kappa, \alpha \beta}^{el}=
   &&
  2 \bigg[ \frac{\Omega_{0}}{(2 \pi)^3}
     \int_{BZ} \sum_{n}^{occ} s \;
        \langle
               u_{n{\bf k}}
               |
             \frac{\partial v_{{\rm ext},{\bf k}}^{\prime} }
                  {\partial \tau_{\kappa \alpha}           }
               |
             \frac{\partial u_{n{\bf k}}       }
                  {\partial {\cal E}_{\beta}   }
        \rangle
       d{\bf k}
+ \frac{1}{2}
   \int_{\Omega_0}
     [\frac {\partial v_{\rm xc0}          }
             {\partial  \tau_{\kappa\alpha} }
       ({\bf r}) ]
     [ \frac {\partial n          }
             {\partial {\cal E}_\beta  }
       ({\bf r}) ]^*
   d{\bf r}     \bigg]
\end{eqnarray}

This equation corresponds to the third formulation of $Z^*$ in which it
appears as the first derivative of the force on the atoms $\kappa$ with
respect to an electric field (Eq. (\ref{EqZ.2})). Indeed, it is 
directly connected to
the following expression of the force, deduced from the Hellmann-Feynman
theorem~:
\begin{eqnarray}
F_{\kappa, \alpha}^{el}=
   &&
   \frac{\Omega_{0}}{(2 \pi)^3}
     \int_{BZ} \sum_{n}^{occ} s \;
        \langle
               u_{n{\bf k}}
               |
             \frac{\partial v_{{\rm ext},{\bf k}}^{\prime} }
                  {\partial \tau_{\kappa \alpha}           }
               |
             u_{n{\bf k}}
        \rangle
       d{\bf k}
+
   \int_{\Omega_0}
     [\frac {\partial v_{\rm xc0}          }
             {\partial  \tau_{\kappa\alpha} }
       ({\bf r}) ]
     [ n
       ({\bf r}) ]^*
   d{\bf r}
\end{eqnarray}

Compared to Eq. (\ref{Z*B}) and Eq.~(41) of Ref.~\cite{Gonze97b}, 
Eq. (\ref{Eq5.9}) has the advantage that the computation of the
first-order wavefunction derivative with respect to the electric field
perturbation is the {\it only} computationally intensive step needed to 
deduce the {\it full} set of effective charges.
We note however that the implementation of Eq. (\ref{Eq5.9}), rather 
easy within a
plane wave -- pseudopotential approach, is not so straightforward 
when the basis
set is dependent on the atomic positions, as in LAPW methods (additional Pulay
terms must be introduced).

\section{Band-by-band decompositions}

\subsection{Displacement of the center of gravity of Wannier functions}

Inspired by a previous discussion by Zak~\cite{Zak89}, Vanderbilt and
King-Smith~\cite{Vanderbilt93} emphasized that the macroscopic electronic
polarization acquires a particular meaning when expressed in terms of localized
Wannier functions. The periodic part of Bloch functions  $u_{n{\bf 
k}}({\bf r})$
are related to the Wannier functions $W_n({\bf r})$ through the following
transformations~:
\begin{eqnarray}
\label{eq.uW}
  u_{n{\bf k}}({\bf r}) &=& \frac{1}{\sqrt{N}}
        \sum_{\bf R} \; e^{-i{\bf k.(r-R)}} \; W_n{\bf (r-R)}
\end{eqnarray}
\begin{eqnarray}
\label{Eq.Wu}
  W_n{\bf (r)} &=& \frac{\sqrt{N} \; \Omega_o}{(2 \pi)^3}
        \int_{BZ} \; e^{i{\bf k.r}} \; u_{n{\bf k}}({\bf r}) \;
d{\bf k}
\end{eqnarray}
{}From this definition, we deduce that~:
\begin{eqnarray}
\frac{\partial}{\partial {\bf k}_{\beta}}  u_{n{\bf k}}({\bf r})
=
\frac{1}{\sqrt{N}} \;
        \sum_{R} \;\; [-i (r_{\beta}-R_{\beta})] \;
                             e^{-i{\bf k.(r-R)}} \; W_n({\bf r-R})
\end{eqnarray}
where {\bf R} runs over all real space lattice vectors. Introducing this
result in Eq. (\ref{Pol}), we obtain~:
\begin{eqnarray}
{\cal P}_{\beta}^{el} = \frac{s}{\Omega_o}
      \sum_{n}^{occ}
        \int  r_{\beta} . | W_n({\bf r}) |^2 \;d{\bf r}
\label{eq.PtW}
\end{eqnarray}
{}From this equation, the electronic part of the polarization is simply
deduced from the position of the center of gravity of the electronic charge
distribution, as expressed in terms of localized Wannier functions. In other
words, for the purpose of determining the polarization, {\it ``the true
quantum mechanical electronic system can be considered as an effective
classical system of quantized point charges, located at the centers of
gravity associated with the occupied Wannier functions in each unit
cell''}~\cite{Vanderbilt93}.

We observe that Eqs. (\ref{eq.uW}) and (\ref{Eq.Wu}) establish a one-to-one
correspondence between $u_{n {\bf k}}$ and $W_n$.  As previously
emphasized in Section III, when working within the {\it diagonal} gauge,
$u_{d,n {\bf k}}$ becomes identified with the Kohn-Sham orbitals so that the
associated $W_{d,n}$ will correspond to a single band Wannier function.
Within this specific gauge, we can therefore isolate $P_{m,\beta}$, the
contribution of band $m$ to the $\beta$ component of the polarization,
by separating the different term in the sum appearing in Eq. (\ref{eq.PtW})~:
\begin{eqnarray}
\label{PWm}
{\cal P}_{m,\beta}^{el}= \frac{s}{\Omega_o}
        \int  r_{\beta} . | W_{d,m}({\bf r}) |^2 d{\bf r}
\end{eqnarray}

If we take the derivative of the polarization with respect to a collective
atomic displacement, $Z^{el}_{\kappa, \alpha \beta}$ can be written in terms
of Wannier functions as~:
\begin{eqnarray}
\label{Z*W}
Z^{el}_{\kappa, \alpha \beta} =
   \sum_{n}^{occ}   s
      \int  r_{\beta}
          [ ({\left . \frac{\partial W_n({\bf r})}{\partial \tau_{\kappa,
\alpha}} \right |}_{d} )^*
              \;  W_{d,n}({\bf r})
          +
            (W_{d,n}({\bf r}))^*
              \;  {\left . \frac{\partial W_n({\bf r})}{\partial \tau_{\kappa,
\alpha}}\right |}_{d}
          ]
       d{\bf r}
\end{eqnarray}
As for the polarization, this equation has a simple physical meaning. In
response to an atomic displacement, the electronic distribution is modified
and the electronic contribution to $Z^*$ can be identified from the
displacement of the center of gravity of the occupied Wannier functions.
Working within the diagonal gauge at any order of perturbation, we will be
able to follow the change of single band Wannier functions all along 
the path of
atomic displacements. In the previous expression,  the contribution of band
$m$ to $Z^{el}_{\kappa, \alpha \beta}$ can be isolated~:
\begin{eqnarray}
\label{Z*Wn}
[Z^{el}_{\kappa, \alpha \beta}]_m =
    s  \int  r_{\beta}
          [ ({\left . \frac{\partial W_m({\bf r})}{\partial \tau_{\kappa,
\alpha}} \right |}_{d})^*
              \;  W_{d,m}({\bf r})
          +
            (W_{d,m}({\bf r}))^*
              \;  {\left . \frac{\partial W_m({\bf r})}{\partial \tau_{\kappa,
\alpha}} \right |}_{d}
          ]
       d{\bf r}
\end{eqnarray}
This equation identifies the contribution from band $m$ to the Born
effective charge as $\Omega_0$ times the {\it change of polarization
corresponding to the displacement of a point charge $s$ on a distance
equal to the displacement of the Wannier center of this band}. Eq.
(\ref{Z*Wn}) can also be estimated from finite difference by combining
Eq. (\ref{EqPf}) and (\ref{PWm}), providing an easy way to decompose
$Z^{el}_{\kappa, \alpha \beta}$ as soon as the Wannier
functions of the system are known \cite{Marzari98}.

Alternatively, Eq.~(\ref{Z*Wn}) can also easily be evaluated in 
reciprocal space~:
\begin{eqnarray}
\label{Z*Bn}
[Z^{el}_{\kappa, \alpha \beta}]_m
=  -2 \frac{\Omega_o}{(2 \pi)^3} \; i \; s
        \int_{BZ}
        \langle
             {\left .
             \frac{\partial u_{m{\bf k}}}{\partial \tau_{\kappa, \alpha}}
             \right |}_{d}
  |
             {\left .
              \frac{\partial u_{m{\bf k}}}{\partial { k}_{\beta}}
             \right |}_{d}
       \rangle
       d{\bf k}
\end{eqnarray}
As Bloch and Wannier functions are related through a band-by-band
transformation, the contribution from band $m$ to $Z^*_{\kappa, \alpha
\beta}$ in Eq. (\ref{Z*Bn}) keeps the same clear physical meaning as in Eq.
(\ref{Z*Wn})~:
\begin{eqnarray}
[Z^{el}_{\kappa, \alpha \beta}]_m = \Omega_0
  .  \Delta{\cal P}^{el}_{m,\beta}= \Omega_0 . s . \Delta d_{\beta}
\end{eqnarray}
where
$\Delta d_{\beta}$ is the displacement in direction $\beta$ of the  Wannier
center of band $m$ induced by the unitary displacement of the sublattice of
atoms $\kappa$ in direction $\alpha$. This decomposition is strictly
equivalent to what is obtained when computing $\Delta{\cal P}^{el}_{m,\beta}$
from finite differences either in real space, using Wannier functions
and Eq. (\ref{PWm}), as reported by Marzari {\it et al.}, or within the Berry
phase approach when separating band by band contribution to Eq.
(\ref{Pol}), the reciprocal space equivalent of Eq. (\ref{PWm}).

In practical calculations, where each band can be thought as a
combinaison of well-known orbitals, the displacement of the Wannier center
is associated to the admixture of a new orbital character to the band 
and must be
attributed to dynamical changes of orbital hybridizations. As illustrated
in some recent
studies~\cite{Massidda95,Ghosez95a,Ghosez95b,Posternak97,Ghosez98}, the
decomposition of $Z^*$ appears therefore as a powerful tool for the microscopic
characterisation of the bonding in solids.

Let us emphasize again that the previous decomposition in terms of a 
single band
is valid only if the {\it diagonal} gauge was used to define the Kohn-Sham
wavefunctions, hence the ``$d$'' subscript in Eq.~(\ref{PWm}), (\ref{Z*Wn}) and
(\ref{Z*Bn}).
The ground-state wavefunctions are conventionally computed within the diagonal
gauge. However, in most calculations,  the first-derivatives of these
wavefunctions are computed within the {\it parallel} gauge. Within this choice,
the change in each Bloch function will be a mixing of different Kohn-Sham
orbitals when the perturbation is applied so that the associated change in
functions $W_n$, defined from Eq. (\ref{Eq.Wu}), will correspond to the change
of a multi-band Wannier function. Evaluating Eq. (\ref{Z*Wn}) or (\ref{Z*Bn})
within such a gauge, we will identify the displacement of a complex of bands
rather than that of a single band. In practice, the first-order derivative of
wavefunctions in
the diagonal gauge $\frac {d u_{n{\bf k}}}{d \lambda} |_d$ can be deduced from
those in the parallel gauge $\frac {d u_{n{\bf k}}}{d \lambda} |_p$ and the
ground-state wavefunctions in the diagonal gauge $u_{d,n{\bf k}}$, by adding
contributions from the subspace of the occupied bands~:
\begin{equation}
\label{Eq.gaugechange}
{\left . \frac {d u_{m{\bf k}}}{d \lambda} \right |}_d  =
{\left . \frac {d u_{m{\bf k}}}{d \lambda} \right |}_p
- \sum_{n \neq m}^{occ}
     \frac{   \langle u_{d,n{\bf k}}
                |  \frac{\partial H} {\partial \lambda}
                |  u_{d,m{\bf k}}
              \rangle
           }
  {   (\epsilon_{n{\bf k}} - \epsilon_{m{\bf k}})
           }
     u_{d,n{\bf k}}
\end{equation}
We note that this transformation (Eq.~(\ref{Eq.gaugechange})) can present some
problems when the denominator vanishes~: this happens when the valence energies
are degenerated. The problem can be partly bypassed by keeping a parallel
transport gauge within the space of degenerated wavefunctions. 
Practically, this
means that we will only be able to separate the contributions of disconnected
set of bands.

\subsection{Other band-by-band decompositions}

After focusing on Eq.~(\ref{Z*B}), we now investigate the possible band-by-band
decompositions of Eq.~(41) of Ref.~\cite{Gonze97b} and Eq.~(\ref{Eq5.9}). 
These expressions, unlike
Eq.~(\ref{Z*B}), are not written as simple sums of matrix elements, 
each related
with a single band. However, individual contributions to Eq.~(\ref{Eq5.9}) can
be identified using the following decomposition of the density~:
\begin{eqnarray}
\label{Eq.n}
n({\bf r})=
   &&
\frac{1}{(2 \pi)^3}
     \int_{BZ} \sum_{n}^{occ} s \;
               u_{n{\bf k}}^*({\bf r})
               u_{n{\bf k}}({\bf r})
       d{\bf k}.
\end{eqnarray}
It gives~:
\begin{eqnarray}
\label{Eq5.9bis}
Z_{\kappa, \alpha \beta}^{el}=
   &&
  2 \frac{\Omega_{0}}{(2 \pi)^3}
     \int_{BZ} \sum_{n}^{occ} s \;
        \langle
               u_{n{\bf k}}
               |
             \frac{\partial v_{{\rm ext},{\bf k}}^{\prime} }
                  {\partial \tau_{\kappa \alpha}           }
              +
				\frac {\partial v_{\rm xc0}         }
             					{\partial 
\tau_{\kappa\alpha} }
               |
             \frac{\partial u_{n{\bf k}}       }
                  {\partial {\cal E}_{\beta}   }
        \rangle
       d{\bf k}
\end{eqnarray}
for which the following decomposition is obtained, using
the diagonal gauge wavefunctions~:
\begin{eqnarray}
\label{Eq5.9Bn}
[\tilde Z_{\kappa, \alpha \beta}^{el}]_m=
   &&
  2 \frac{\Omega_{0}}{(2 \pi)^3}
     \int_{BZ} s \;
        \langle
               u_{d,m{\bf k}}
               |
             \frac{\partial v_{{\rm ext},{\bf k}}^{\prime} }
                  {\partial \tau_{\kappa \alpha}           }
              +
				\frac {\partial v_{\rm xc0}         }
             					{\partial 
\tau_{\kappa\alpha} }
               |
             {\left .  \frac{\partial u_{m{\bf k}}       }
                  {\partial {\cal E}_{\beta}   }  \right |}_d
        \rangle
       d{\bf k}
\end{eqnarray}
This expression corresponds to the contribution of the electrons of
band $m$ to the force induced on atom $\kappa$ by a macroscopic field
${\cal E}_{\beta}$. However, it is not equivalent to Eqs.~(\ref{Z*Wn})
or ~(\ref{Z*Bn}). Indeed, for a particular band
$m$, the difference between matrix elements present in Eq.~(\ref{Z*Bn})
and~(\ref{Eq5.9Bn}) is (within a given gauge)~:
\begin{eqnarray}
\label{Eq.diff}
&& \hspace{-20mm}
\bigg[
    \langle
               u_{m{\bf k}}
               |
             \frac{\partial v_{{\rm ext},{\bf k}}^{\prime} }
                  {\partial \tau_{\kappa \alpha}           }
       +
				\frac {\partial v_{\rm xc0}         }
             					{\partial 
\tau_{\kappa\alpha} }
               |
             \frac{\partial u_{m{\bf k}}       }
                  {\partial {\cal E}_{\beta}   }
    \rangle
\bigg]
-
\bigg[
      \langle
               \frac{ \partial u_{m{\bf k}}           }
                    { \partial \tau_{\kappa, \alpha}  }
               |
               -i \frac{ \partial u_{m{\bf k}}    }
                       { \partial {k}_{\beta}     }
      \rangle
\bigg] =
\nonumber \\
& & \hspace{15mm}
-
  \frac{1}{2}  \int_{\Omega_0}
       K_{\rm xc}({\bf r,r'})
       [ \frac{ \partial {n}                 }
              { \partial \tau_{\kappa \alpha}}
         ({\bf r}) ]^*
         \frac{ \partial n_{m{\bf k}}     }
              { \partial {\cal E}_\beta   }
         ({\bf r'} )
        d{\bf r}  \; d{\bf r'}
\nonumber \\
  & &  \hspace{15mm}
+
  \frac{1}{2}  \int_{\Omega_0}
       K_{\rm xc}({\bf r,r'})
       [ \frac{ \partial n_{m{\bf k}}         }
              { \partial \tau_{\kappa \alpha} }
         ({\bf r})
                  ]^*
         \frac{ \partial n                    }
              { \partial {\cal E}_\beta       }
         ({\bf r'} )
        d{\bf r}  \; d{\bf r'}
\end{eqnarray}
where $n_{m{\bf k}}({\bf r})$ is a short notation for $u_{m{\bf k}}^*({\bf r})
u_{m{\bf k}}({\bf r})$. The summation of these differences on all the bands and
integration on the Brillouin zone gives zero, as expected. However, the
band-by-band difference, Eq.~(\ref{Eq.diff}), does not vanish. This 
demonstrates
that the quantity defined from Eq. (\ref{Eq5.9Bn}) is independent from that of
Eq.~(\ref{Z*Wn}) and has therefore no specific meaning in terms of Wannier
functions. Unlike Eq.~(\ref{Eq5.9}), Eq.~(41) of Ref.~\cite{Gonze97b} is 
not naturally convertible in a sum of independent band contributions.

\section{Numerical comparison}

The previous theoretical results can now be illustrated on a numerical
example. In what follows, we will consider the case of barium titanate
(BaTiO$_{3}$), a well-known ferroelectric material which is stable at
high temperature in a cubic perovskite structure and exhibits non-trivial
values of $Z^*$~\cite{Ghosez98}.

Our calculations have been performed within the density functional
theory and the local density approximation~\cite{Jones89}.  For the
exchange-correlation energy, we used a polynomial
parametrization~\cite{Goedeker96} of Ceperley-Alder~\cite{Ceperley80}
homogeneous electron gas data. We adopted a
planewave-pseudopotential approach. We choose highly transferable
extended norm-conserving pseudopotentials as described in Ref.
\cite{Teter93}.
The Ba 5s, Ba 5p, Ba 6s, Ti 3s, Ti 3p, Ti 3d, Ti 4p, O 2s, and O2p
levels have been treated has valence states. The electronic wavefunction has
been expanded in plane-waves up to a kinetic energy cutoff of 35 hartrees.
Integrals over the Brillouin zone have been replaced by sums on a
$6 \times 6 \times 6$ mesh of special k-points. The Born effective charges
have been computed in the cubic phase at the optimized lattice parameter
of 3.94 \AA. They have been obtained by linear response following the
scheme described in Ref.~\cite{Gonze97b}.

In Table~\ref{Zbbb}, we summarize the results obtained from independent
formulations for the titanium charge ($Z^*_{Ti}$). The decomposition of
total $Z^{*}$ is provided, according to Eq.~(\ref{Z*Bn}) and 
Eq.~(\ref{Eq5.9Bn})
in the diagonal gauge and to Eq.~(\ref{Z*Bn}) in the parallel gauge. We
also compare our results to those reported independently by Marzari
{\it et al.}~\cite{Marzari98} from a direct computation of the
displacement of the center of gravity of the electronic Wannier
functions.

As expected in this class of compounds~\cite{Ghosez98}, the total charge
on the Ti atom is anomalously large ($+7.25$) and comparable in amplitude
to the value of $+7.16$ reported independently using the Berry phase
approach~\cite{Zhong94}. The main anomalous contribution (deviation from
the nominal value of the second column) is located in the O 2p bands. 
Similarly, the oxygen charge along the Ti--O bond is anomalously large
and equal to $-5.71$. Both Ti and O anomalous charge 
contributions are related to each other and can be assigned to dynamical 
changes of hybridization between O 2p and Ti 3d orbitals~\cite{Ghosez98}. 
This was explicitely demonstrated for a parent compound (KNbO$_3$) by 
Posternak {\it et al.}~\cite{Posternak94}.

We observe that the global charge is equivalent independently of the
approach while it is not the case for partial contributions coming
from different isolated sets of bands. First, the
band-by-band decompositions obtained within the diagonal and parallel gauges
are not similar. This means that the unitary transform performed when
changing the gauge strongly mix the different bands. Second, the results
deduced from Eq.~(\ref{Z*Bn}) and ~(\ref{Eq5.9Bn}) within the diagonal
gauge are {\it significantly} different, demonstrating that the amplitude of
the quantity defined in Eq.~(\ref{Eq.diff}) is not negligible. Third, the
results  obtained from Eq.~(\ref{Z*Bn}) within the diagonal gauge are 
comparable
to those of Marzari {\it et al.} who explicitely computed the 
electronic Wannier
functions and estimated Eq.~(\ref{Z*Wn}) using a finite difference
technique combining Eq.~(\ref{EqPf}) and Eq.~(\ref{PWm}). This
illustrates the physical interpretation of Eq.~(\ref{Z*Bn}) in terms
of localized Wannier functions~: the contributions describe the displacement
of the Wannier center of each given set of bands, induced in response to the
displacement of the Ti atom.

\section{Conclusions}

In conclusion, the Born effective charges can be computed from
first-principles using different techniques and algorithms. The global
charge is a gauge invariant quantity and is obtained independently of
the approach while a special care is needed to separate individual
contributions from separate groups of occupied bands. When using
linear response techniques, the identification of band-by-band
contributions, equivalent to those obtained within the Berry phase
approach, requires the use of Eq.~(\ref{Z*Bn}), when working within the
diagonal gauge. The contribution $[Z^{el}_{\kappa, \alpha \beta}]_m$
is then directly related to the displacement in direction $\beta$ of
the Wannier center of band $m$, when displacing the sublattice of atoms
$\kappa$ in direction $\alpha$. The diagonal gauge condition is mandatory
to identify single band contributions. The results obtained are
conceptually and numerically different from those computed when using
Eq. (\ref{Eq5.9Bn}), independently of the gauge choice.

\section{Acknowledgments}
We thank N. Marzari for interesting discussions. X.G. acknowledges
financial support from the F.N.R.S (Belgium), from the program ``Pole
d'attraction interuniversitaire'' P4/10 of the Belgian SSTC,
and from the FNRS-FRFC project 2.4556.99.

\newpage
\begin{table}
\caption {Band-by-band decompositions of the Born effective charge of
the Ti atom in the cubic phase of BaTiO$_3$. The first line refer to
the pseudo-ion charge while the other contributions comes from the
different valence electron levels. The nominal values expected in a
purely ionic material are reported in the second column;
band-by-band contributions presented in the three next columns were
computed from linear response first-principles calculations. The last
column refer to first-principles values deduced from the computation of
Wannier functions~\protect\cite{Marzari98}.}
\label{Zbbb}
\begin{center}
\begin{tabular}{l|c|ccc|c}
&Reference  &\multicolumn{3}{c|}{Linear response}  & Wannier functions \\
& nominal &\multicolumn{2}{c}{Diagonal gauge} &Parallel gauge & 
Diagonal Gauge \\
& charges  & from Eq. (\ref{Z*Bn}) & from Eq. (\ref{Eq5.9Bn})
& from Eq. (\ref{Z*Bn}) & from Eq. (\ref{Z*Wn})\\
\hline
$Z_{Ti}$ &$+12.00$   &$+12.00$   &$+12.00$  &$+12.00$  &+12.00\\
Ti 3s    &$-2.00$     &$-2.03$   &$+1.56$   &$-0.36$  &$-2.04$\\
Ti 3p    &$-6.00$     &$-6.22$   &$-9.54$   &$-5.50$  &$-6.19$\\
Ba 5s    &$0.00$      &$+0.05$   &$-0.36$   &$0.00$   &$+0.04$\\
O 2s      &$0.00$     &$+0.23$   &$-1.56$   &$-0.41$   &$+0.20$\\
Ba 5p    &$0.00$     &$+0.36$   &$+1.47$   &$+0.10$   &$+0.31$\\
O 2p     &$0.00$     &$+2.86$   &$+3.68$   &$+1.42$   &$+3.01$\\
\hline
$Z^*_{Ti}$  &$+4.00$    &$+7.25$    &$+7.25$  &$+7.25$  &$+7.33$ \\
\end{tabular}
\end{center}
\end{table}


\begin{thebibliography}{10}
\bibitem{Notation}
This lengthy expression will not be reproduced here.
\end{thebibliography}
\end{document}